\documentclass[12pt]{article}
\usepackage{amssymb, amsmath, graphicx}

\def\R{\mathbb{R}}

\newtheorem{theorem}{Theorem}

\newtheorem{lemma}[theorem]{Lemma}

\begin{document}

\author{\textbf{Martin Grothaus} \\
{\small Mathematics Department, University of Kaiserslautern,}\\
{\small D 67653 Kaiserslautern, Germany}\\
{\small grothaus@mathematik.uni-kl.de} \and \textbf{Maria Jo\~{a}o Oliveira} 
\\
{\small Universidade Aberta, P 1269-001 Lisbon, Portugal}\\
{\small CMAF, University of Lisbon, P 1649-003 Lisbon, Portugal}\\
{\small oliveira@cii.fc.ul.pt} \and \textbf{Jos\'e Lu{\'\i}s da Silva} \\
{\small DME, University of Madeira, P 9000-390 Funchal, Portugal}\\
{\small CCM, University of Madeira, P 9000-390 Funchal, Portugal}\\
{\small luis@uma.pt} \and \textbf{Ludwig Streit} \\
{\small Forschungszentrum BiBoS, Bielefeld University, D 33501 Bielefeld,
Germany}\\
{\small CCM, University of Madeira, P 9000-390 Funchal, Portugal}\\
{\small streit@physik.uni-bielefeld.de}}
\title{Self-avoiding fractional Brownian motion - The Edwards model}
\date{}
\maketitle

\begin{abstract}
In this work we extend Varadhan's construction of the Edwards polymer
model to the case of fractional Brownian motions in $\R^d$, for any dimension 
$d\geq 2$, with arbitrary Hurst parameters $H\leq 1/d$.
\end{abstract}


\medskip

\noindent
\textbf{2000 AMS Classification:} 60G15, 60G18, 60J55, 28C20, 46F25

\vspace*{-0.5cm}

\section{Introduction}

In recent years the fractional Brownian motion has become an object of
intense study  due to its special properties, such as short/long range
dependence and self-similarity, leading to proper and natural  applications
in different fields. In particular, the specific properties  of fractional
Brownian motion paths have been used e.g.~in the modelling of polymers. For 
the self-intersection properties of sample paths see e.g.~\cite{GRV03}, 
\cite{NH05}, \cite{NH07}, \cite{HNS08}, \cite{R87}, and for the intersection 
properties with other independent fractional Brownian motion see 
e.g.~\cite{NL07}, \cite{OSS09} and references therein. Comments on the 
relevance of fractional Brownian motion  for polymer modelling, in particular 
with $H=1/3$ for polymers in a compact or collapsed phase, can e.g.~be found 
in \cite{Biswas}.

The fractional Brownian motion on $\mathbb{R}^{d}$, $d\geq 1$, with Hurst
parameter $H\in \left( 0,1\right) $ is a $d$-dimensional centered Gaussian
process $B^{H}=\{B_{t}^{H}:t\geq 0\}$ with covariance function 
\begin{equation*}
\mathbb{E}(B_{t}^{H,i}B_{s}^{H,j})=\frac{\delta _{ij}}{2}\left(
t^{2H}+s^{2H}-|t-s|^{2H}\right) ,\quad i,j=1,\ldots ,d,\ s,t\geq 0.
\end{equation*}
An informal but suggestive definition of self-intersection local time of a
fractional Brownian motion $B^{H}$ is given in terms of an integral over a
Dirac $\delta $-function 
\begin{equation*}
L=\int_{0}^{T}dt\int_{0}^{T}ds\,\delta (B^{H}(t)-B^{H}(s)),
\end{equation*}
intended to measure the amount of time the process spends intersecting
itself in a time interval $\left[ 0,T\right] $. A rigorous definition may be
given by approximating the $\delta $-function by the heat kernel 
\begin{equation*}
p_{\varepsilon }(x):=\frac{1}{(2\pi \varepsilon )^{d/2}}e^{-\frac{|x|^{2}}{%
2\varepsilon }},\quad x\in\R^d, \varepsilon >0,
\end{equation*}
which leads to the approximated self-intersection local time 
\begin{equation}
L_{\varepsilon }:=\int_{0}^{T}dt\int_{0}^{t}ds\,p_{\varepsilon
}(B^{H}(t)-B^{H}(s)).  \label{3Eq2}
\end{equation}
The main problem is then the removal of the approximation, that is, 
$\varepsilon \searrow 0$.

In the classic Brownian motion case ($H=1/2$), $L_{\varepsilon }$ converges
in $L^{2}$ only for $d=1$. To ensure the existence of a limiting process for
higher dimensions one must center the approximated self-intersection 
\begin{equation}
L_{\varepsilon, c}:= L_{\varepsilon }-\mathbb{E}(L_{\varepsilon }).  \label{3Eq1}
\end{equation}
For the case of the planar Brownian motion this is sufficient to ensure the $%
L^{2}$-convergence of (\ref{3Eq1}) as $\varepsilon $ tends to zero \cite{V69}%
, but for $d\geq 3$ a further multiplicative renormalization $r(\varepsilon )
$ is required to yield a limiting process, now as a limit in law of 
\begin{equation}
r(\varepsilon )\left( L_{\varepsilon }-\mathbb{E}(L_{\varepsilon })\right) .
\label{3Eq3}
\end{equation}
Through a different approximation, this has been shown in \cite{CY87}, \cite%
{Y85}.

Extending  Varadhan's results to the planar fractional Brownian motion,
Rosen in \cite{R87} shows that, for $1/2<H<3/4$, the centered approximated
self-intersection local time converges in $L^{2}$ as $\varepsilon $ tends to
zero.

This result, as well as all the above quoted ones for the classic Brownian
motion, have been extended by Hu and Nualart in \cite{NH05} to any $d$%
-dimensional fractional Brownian motion with $H<3/4$. More precisely, Hu and
Nualart have shown that for $H<1/d$ the approximated self-intersection local
time (\ref{3Eq2}) always converges in $L^{2}$. For $1/d\leq H<3/(2d)$, a $%
L^{2}$-convergence result still holds, but now for the centered approximated
self-intersection local time (\ref{3Eq1}). In this case, 
\begin{equation}
\mathbb{E}(L_{\varepsilon })=\left\{ 
\begin{array}{cl}
& TC_{H,d}\varepsilon ^{-d/2+1/(2H)}+o(\varepsilon ),\quad \mathrm{if}\
1/d<H<3/(2d)\\ 
& \\ 
& \frac{T}{2H(2\pi )^{d/2}}\ln (1/\varepsilon )+o(\varepsilon ),\quad 
\mathrm{if}\ H=1/d
\end{array}
\right. ,\label{3Eq4} 
\end{equation}
where $C_{H,d}$ is a positive constant which depends of $H$ and $d$. In 
particular, for $1/d\leq H<\min\{3/(2d),2/(d+1)\}$, an explicit integral 
representation for the mean square limiting process $L_c$ as an It\^o integral 
is even obtained in \cite{HNS08}. For $3/(2d)\leq H<3/4$, a multiplicative 
renormalization factor $r(\varepsilon )$ is required in \cite{NH05} to prove 
the convergence in distribution of the random variable (\ref{3Eq3}) to a 
normal law as $\varepsilon $ tends to zero.

To model polymers by Brownian paths Edwards \cite{E65} proposed to suppress
self-intersections by a factor
\begin{equation*}
\exp \left( -gL\right) ,
\end{equation*}
with $g>0.$ For planar Brownian motion Varadhan \cite{V69} showed that the 
expectation value $\mathbb{E}(L_\varepsilon)$ has a logarithmic divergence but 
after its subtraction the centered $L_{\varepsilon, c}$ converges in $L^2$, 
with a suitable rate of convergence. From this, Varadhan could conclude the 
integrability of $\exp(-gL_c)$, thus giving a proper meaning to the Edwards 
model. For more details see also \cite{Si74}. In the three-dimensional case 
this is clearly much more difficult \cite{Bolthausen}, \cite{W80}.

In this note we extend Varadhan's construction to arbitrary spatial
dimension $d\geq 2$ and Hurst parameters $H\leq 1/d$. For this, the 
convergence results proved in \cite{NH05} will be essential. Because of this,
in the following section we collect from \cite{NH05} the necessary
information on fractional Brownian motion and its self-intersection local
time, and in Section 3 we state and prove the existence theorem (Theorem 
\ref{Theorem2}).

\section{Preliminaries}

As shown in \cite{NH05}, given a $d$-dimensional fractional Brownian motion $%
B^H$ with Hurst parameter $H\in\left(0,1\right)$, for each $\varepsilon>0$
the approximated self-intersection local time (\ref{3Eq2}) is a square
integrable random variable with 
\begin{equation*}
\mathbb{E}(L_\varepsilon^2)=\frac{1}{(2\pi)^d} \int_{\mathcal{T}}d\tau \,%
\frac{1}{((\lambda +\varepsilon )(\rho +\varepsilon)-\mu ^{2})^{d/2}}, 
\end{equation*}
where 
\begin{equation*}
\mathcal{T}:=\{(s,t,s^{\prime },t^{\prime }): 0<s<t<T,0<s^{\prime
}<t^{\prime }<T\}  
\end{equation*}
and for each $\tau=(s,t,s^{\prime },t^{\prime })\in\mathcal{T}$, 
\begin{equation}
\lambda(\tau):=(t-s)^{2H},\quad \rho(\tau):=(t^{\prime }-s^{\prime})^{2H}, 
\label{6Eq1}
\end{equation}
and 
\begin{equation}
\mu(\tau):=\frac{1}{2}\left[|s-t^{\prime}|^{2H}+|s^{\prime 2H}-t|^{2H}-|t-t^{\prime}|^{2H}-|s-s^{\prime}|^{2H}\right]. \label{6Eq2}
\end{equation}
Furthermore, for each $\varepsilon, \gamma > 0$ is 
\begin{align}
&\mathbb{E}(L_{\varepsilon }L_{\gamma})-\mathbb{E} (L_{\varepsilon})
\mathbb{E}(L_{\gamma})=\label{esp} \\
&\frac{1}{(2\pi )^{d}}\int_{\mathcal{T}}d\tau \,\left( \frac{1}{((\lambda
+\varepsilon )(\rho +\gamma )-\mu ^{2})^{d/2}}-\frac{1}{((\lambda +\varepsilon
)(\rho +\gamma))^{d/2}}\right) =:E_{\varepsilon\gamma}. \label{eqE1} 
\end{align}
Note that the integral in \eqref{eqE1} is also well-defined for all 
$\varepsilon, \gamma \ge 0$ (however it might be infinite). Hence, using this 
integral representation, we can extend $E_{\varepsilon\gamma}$ to general 
$\varepsilon, \gamma \ge 0$. This is contrast with \eqref{esp} 
which in general is not well-defined for $\varepsilon = 0$ and/or 
$\gamma = 0$.  

From \eqref{eqE1} one can easily derive that a necessary and sufficient 
condition for convergence of 
$L_{\varepsilon, c} = L_\varepsilon - \mathbb{E}(L_\varepsilon)$ to a limiting
process $L_c$ in $L^2$ as $\varepsilon \searrow 0$ is that $E_{00}<\infty$. As
shown in \cite[Lemma 11]{NH05}, the integral $E_{00}$ is finite if and only
if $dH<3/2$.

\section{Theorems and Proofs}

\begin{theorem}\label{Theorem1} Assume that $(d+1)H < 3/2$, $d \ge 2$. Then 
there exists a positive constant $K$ such that 
\begin{equation*}
\mathbb{E}\left( \left(L_{\varepsilon, c } - L_c\right)^{2}\right) 
\leq K \varepsilon^{1/2}  
\end{equation*}
for all $\varepsilon >0$.
\end{theorem}
\noindent \textbf{Proof.} Using \eqref{eqE1}, a simple calculation and taking 
the limit $\gamma \searrow 0$ yields 
\begin{equation*}
\mathbb{E}\left(\left(L_{\varepsilon, c} - L_c \right)^2\right) = (E_{\varepsilon\varepsilon}-
E_{\varepsilon0})+(E_{00}-E_{\varepsilon0}) 
\end{equation*}
with 
\begin{multline*}
E_{\varepsilon \varepsilon}-E_{\varepsilon 0}
= \frac{d}{2(2\pi )^{d}}\int_{\mathcal{T}}d\tau\, (\lambda +\varepsilon)
\int_{0}^{\varepsilon }dx \\
\left( \frac{1 }{\left( (\lambda
+\varepsilon)(\rho+x) \right) ^{d/2+1}}-\frac{1}{\left((\lambda
+\varepsilon)(\rho +x)-\mu^2\right)^{d/2+1}}\right) \le 0.
\end{multline*}
Hence
\begin{multline}\label{diff}
\mathbb{E}\left(\left(L_{\varepsilon, c} - L_c \right)^2\right) \le
E_{00}-E_{\varepsilon 0} \\ = \frac{d}{2(2\pi )^{d}}\int_{\mathcal{T}}d\tau\,
\rho \int_{0}^{\varepsilon }dx\,\left( \frac{1}{(\delta +x\rho )^{d/2+1}}-%
\frac{1}{\left( (\lambda +x)\rho \right) ^{d/2+1}}\right),
\end{multline}
where $\delta:=\lambda\rho-\mu^2$. Thus it is sufficient to establish a suitable upper bound
for \eqref{diff}. Technically, this will follow closely the proof of Lemma 11 in \cite{NH05},
based on the decomposition of the region $\mathcal{T}$ into three subregions
\begin{equation*}
\mathcal{T}\cap\{s<s^{\prime }\}=\mathcal{T}_1\cup\mathcal{T}_2\cup\mathcal{T%
}_3, 
\end{equation*}
where 
\begin{eqnarray*}
&&\mathcal{T}_1:=\{(t,s,t^{\prime },s^{\prime }): 0<s<s^{\prime
}<t<t^{\prime }<T\}, \\
&&\mathcal{T}_2:=\{(t,s,t^{\prime },s^{\prime }): 0<s<s^{\prime }<t^{\prime
}<t<T\}, \\
&&\mathcal{T}_3:=\{(t,s,t^{\prime },s^{\prime }): 0<s<t<s^{\prime
}<t^{\prime }<T\}.
\end{eqnarray*}
Each substitution of $\mathcal{T}$ in (\ref{diff}) by a subregion $\mathcal{T%
}_i$, $i=1,2,3$, yields a different case and for each particular case we will
then establish a suitable upper bound.

As in \cite{NH05}, we will denote by $k$ a generic positive constant which may 
be different from one expression to another one. We set $D := d+1$.

\medskip

\noindent
{\bf Subregion $\mathcal{T}_1$:} We do the change of variables 
$a:=s^{\prime }-s$, $b:=t-s^{\prime }$, and $c=t^{\prime }-t$ for 
$(t,s,t^{\prime },s^{\prime })\in \mathcal{T}_1$. Thus, on $\mathcal{T}_1$, for 
the functions $\lambda$, $\rho$, and $\mu$ defined in \eqref{6Eq1} and 
\eqref{6Eq2} we have
\begin{eqnarray*} 
&&\lambda(t,s,t^{\prime },s^{\prime })=:\lambda_1(a,b,c)= (a+b)^{2H},\quad 
\rho(t,s,t^{\prime },s^{\prime })=:\rho_1(a,b,c)=(b+c)^{2H}\\
&&\mu(t,s,t^{\prime },s^{\prime })=:\mu_1(a,b,c)=\frac{1}{2}\left[(a+b+c)^{2H} + b^{2H} - c^{2H} -  a^{2H}\right].
\end{eqnarray*}
On the region $\mathcal{T}_1$ one can bound (\ref{diff}) by the first term 
only, and to estimate the latter we shall use Lemma \ref{streit} below, 
yielding 
\begin{equation*}
\rho_1 \int_{0}^{\varepsilon }dx\,\frac{1}{(\delta_1 +x\rho_1 )^{(D+1)/2}}\leq
A\varepsilon ^{1/2}\rho_1^{1/2} \delta_1^{-D/2}.
\end{equation*}
From \cite[eq.~(59)]{NH05},
\begin{equation*}
\delta_1 \ge k (a+b)^H(b+c)^Ha^Hc^H \ge k (abc)^{4H/3},
\end{equation*}
we deduce 
\begin{equation*}
\int_{[0,T]^3} da\,db\,dc\, \delta_1^{-D/2} \le k \int_{[0,T]^3} da\,db\,dc\, (abc)^{-2DH/3} < \infty,
\end{equation*}
because $DH < 3/2$. In conclusion
the part of (\ref{diff}) stemming from integration over $\mathcal{T}_{1}$ is
of order $\varepsilon ^{1/2}$.

On the subregions $\mathcal{T}_i$, $i=2,3$, we have to consider the
difference
\begin{equation*}
\Xi_i^\varepsilon := \rho_i \int_{0}^{\varepsilon }dx\left( \frac{1}{(\delta_i +x\rho_i )^{(D+1)/2}}
-\frac{1}{\left( (\lambda_i +x)\rho_i \right) ^{(D+1)/2}}\right), \quad \varepsilon > 0.
\end{equation*}

\medskip

\noindent
{\bf Subregion $\mathcal{T}_2$:} In this case we do the change of variables 
$a:=s^{\prime }-s$, $b:=t^{\prime }-s^{\prime }$, and $c=t- t^{\prime }$ for 
$(t,s,t^{\prime },s^{\prime })\in \mathcal{T}_2$. That is, on $\mathcal{T}_2$
we will have 
\begin{eqnarray*} 
&&\lambda(t,s,t^{\prime },s^{\prime })=:\lambda_2(a,b,c)=b^{2H},\quad 
\rho(t,s,t^{\prime },s^{\prime })=:\rho_2(a,b,c)=(a+b+c)^{2H}\\
&&\mu(t,s,t^{\prime },s^{\prime })=:\mu_2(a,b,c)=\frac{1}{2}\left[(b+c)^{2H} + (a+ b)^{2H}- c^{2H} -  a^{2H}\right].
\end{eqnarray*}
In this case we decompose the corresponding integral \eqref{diff} over the 
regions $\{b \ge \eta a\}$, $\{b \ge \eta c\}$, and 
$\{b < \eta a, b < \eta c\}$, for some fixed but arbitrary $\eta > 0$. We have 
by \eqref{356}, see Appendix, 
\begin{eqnarray*}
\int_{b \ge \eta a} da\,db\,dc\, \Xi_2^\varepsilon &\le& C \varepsilon^{1/2}
\int_{b \ge \eta a} da\,db\,dc\, \rho_2^{1/2} \left( \lambda_2 \rho_2 \right) ^{-D/2} \\
&\le& k \varepsilon^{1/2} \int_{b \ge \eta a} \frac{da\,db\,dc}{(a+b+c)^{DH} b^{DH}}.
\end{eqnarray*}
If $DH < 1$, the integral is finite. If $1 < DH < 3/2$, then by Young 
inequality
\begin{eqnarray*}
\int_{b \ge \eta a} da\,db\,dc\, \Xi_2^\varepsilon &\le& k \varepsilon^{1/2}
\int_0^T\int_0^T \frac{da\,dc}{(a+c)^{DH}} \int_{\eta a}^T db\,b^{-DH} \\
&\le& k \varepsilon^{1/2} \int_0^Tda\,a^{-4DH/3 +1}\int_0^T dc\, c^{-2DH/3} < \infty.
\end{eqnarray*}
In the case $DH = 1$ we have
\begin{equation*}
\int_{b \ge \eta a} da\,db\,dc\, \Xi_2^\varepsilon
\le k \varepsilon^{1/2} \int_0^T dc\,c^{-2/3}\int_0^T da\, a^{-1/3}\ln(T/(\eta a)) < \infty.
\end{equation*}
The case $b \ge \eta c$ can be treated analogously.

To handle the case $b < \eta a$ and $b < \eta c$ we first observe that 
\begin{eqnarray*}
\mu_2 &=& \frac{1}{2}\left(a^{2H}\left(\left(1 + \frac{b}{a}\right)^{2H} - 1 \right)
+ c^{2H}\left(\left(1 + \frac{b}{c}\right)^{2H} - 1 \right)\right) \\ 
&\le& k \left(a^{2H-1} + c^{2H-1}\right)b
\end{eqnarray*}
for sufficiently small $\eta > 0$. Hence, together with \eqref{357}, see Appendix, we obtain
\begin{multline*}
\int_{b < \eta a, b < \eta c} da\,db\,dc\, \Xi_2^\varepsilon \le 
C \varepsilon^{1/2} \int_{b < \eta a, b < \eta c} da\,db\,dc\,
\rho_2^{1/2} \mu_2^{2} \left(\lambda_2\rho_2\right) ^{-(D+2)/2} \\
\le k \varepsilon^{1/2} \int_{b < \eta a, b < \eta c} da\,db\,dc\,
\left(a^{4H-2} + c^{4H-2}\right)(a+b+c)^{-2H-DH}b^{2-2H-DH} \\
\le k \varepsilon^{1/2} \int_{b < \eta a, b < \eta c} da\,db\,dc\,
b^{-DH} (a+b+c)^{-2H-DH} \\ \times \left(a^{(2-D/3)H}b^{DH/3} + c^{(2-D/3)H}b^{DH/3}\right) \\
\le k \varepsilon^{1/2} \int_{[0,T]^3} da\,db\,dc\,
b^{-DH} (a+b+c)^{-2H-DH} a^{(2-D/3)H}b^{DH/3} \\
\le k \varepsilon^{1/2} \int_{[0,T]^3} da\,db\,dc\,
b^{-2DH/3} c^{-2DH/3} a^{-2DH/3} < \infty,
\end{multline*}
because $DH < 3/2$.

\medskip

\noindent
{\bf Subregion $\mathcal{T}_3$:} We do the change of variables $a:=t-s$, 
$b:=s^{\prime }-t$, and $c=t^{\prime }-s^{\prime }$ for 
$(t,s,t^{\prime },s^{\prime })\in \mathcal{T}_3$ . Thus, on $\mathcal{T}_3$, we
have
\begin{eqnarray*} 
&&\lambda(t,s,t^{\prime },s^{\prime })=:\lambda_3(a,b,c)=a^{2H},\quad \rho(t,s,t^{\prime },s^{\prime })=:\rho_3(a,b,c)=c^{2H}\\
&&\mu(t,s,t^{\prime },s^{\prime })=:\mu_3(a,b,c)=\frac{1}{2}\left[(a+b+c)^{2H} + b^{2H} - (b+c)^{2H}- (a+b)^{2H}\right].
\end{eqnarray*}
In this case we decompose the corresponding integral \eqref{diff} over the
regions $\{a \ge \eta_1 b, c \ge \eta_2 b\}$, $\{a < \eta_1 b, c < \eta_2 b\}$,
$\{a \ge \eta_1 b, c < \eta_2 b\}$, and $\{a < \eta_1 b, c \ge \eta_2 b\}$
for some fixed but arbitrary $\eta_1, \eta_2 > 0$. By symmetry it suffices to 
consider the first three regions. Using \eqref{356}, see Appendix, we obtain 
\begin{multline*}
\int_{a \ge \eta_1 b, c \ge \eta_2 b} da\,db\,dc\, \Xi_3^\varepsilon \le C \varepsilon^{1/2}
\int_{a \ge \eta_1 b, c \ge \eta_2 b} da\,db\,dc\, \rho_3^{1/2} \left( \lambda_3 \rho_3 \right) ^{-D/2} \\
\le k \varepsilon^{1/2} \int_0^T db \int_{\eta_1b}^T \frac{da}{a^{DH}} \int_{\eta_2b}^T \frac{dc}{c^{DH}} 
\le k \varepsilon^{1/2} \int_0^T \frac{db}{b^{2DH-2}} < \infty.
\end{multline*}
For the region $\{a < \eta_1 b, c < \eta_2 b\}$, we observe that since 
$H < 3/(2D) \le 1/2$, we can conclude from \eqref{357}, see Appendix,
together with \cite[eq.~(55)]{NH05}, i.e., $\mu_3\leq kb^{2H-2}ac$, that
\begin{eqnarray*}
\Xi_3^\varepsilon &\le& C \varepsilon^{1/2} \rho_3^{1/2} \mu_3^{2} \left(\lambda_3\rho_3\right) ^{-(D+2)/2} \\
&\le& k \varepsilon^{1/2} b^{4H-4}a^{2-2H-DH}c^{2-2H-DH}
\le k \varepsilon^{1/2} a^{-2DH/3}c^{-2DH/3}b^{-2DH/3},
\end{eqnarray*}
which is integrable. Finally, we consider the case
$\{a \ge \eta_1 b, c < \eta_2 b\}$. For $\eta_2 > 0$ small enough we have 
\begin{eqnarray*}
\mu_3 &=& \frac{1}{2}\left((a+b)^{2H}\left(\left(1 + \frac{c}{a+b}\right)^{2H} - 1 \right)
- b^{2H}\left(\left(1 + \frac{c}{b}\right)^{2H} - 1 \right)\right) \\ 
&\le& k \left((a+b)^{2H-1} + b^{2H-1}\right)c
= k b^{2H-1}\left(\left(1 + \frac{a}{b}\right)^{2H-1} + 1 \right)c \le k b^{2H-1}c,
\end{eqnarray*}
where in the last estimate we used $2H-1 < 0$ (due to $H < 1/2$). Then using 
\eqref{357} we obtain
\begin{multline*}
\int_{a \ge \eta_1 b, c < \eta_2 b} da\,db\,dc\, \Xi_3^\varepsilon \le C \varepsilon^{1/2}
\int_{a \ge \eta_1 b, c < \eta_2 b} da\,db\,dc\, \rho_3^{1/2} \mu_3^{2} \left(\lambda_3\rho_3\right) ^{-(D+2)/2}  \\
\le k \varepsilon^{1/2} \int_{a \ge \eta_1 b} da\,db\,
b^{4H-2} a^{-2H-DH} \int_0^{\eta_2 b} dc\, c^{2-2H-DH} \\
\le k \varepsilon^{1/2} \int_0^T da \,a^{-2H-DH} \int_0^{a/\eta_2} db\, b^{-DH+2H+1} 
\le k \varepsilon^{1/2} \int_0^T da \,a^{-2DH+2},
\end{multline*}
which is finite because $DH < 3/2$.
\hfill $\blacksquare \medskip$

\begin{theorem}\label{Theorem2} (i) Assume that $dH=1$, $d \ge 2$. Then there 
exists a positive constant $M$ such that for all $0 \le g \le M$   
\begin{equation}\label{l1}
\exp(- gL_c)
\end{equation}
is an integrable function.\newline
(ii) Assume that $dH < 1$, $d \ge 2$. Then there exists
\[
L := \lim_{\varepsilon \searrow 0} L_{\varepsilon } \ \mbox{in} \ L^2
\]
and for all non-negative constants $g$
\begin{equation*}
\exp(- gL)
\end{equation*}
is an integrable function.
\end{theorem}
\noindent \textbf{Proof.} (i) The case $d=2$ and $H=1/2$ was treated in 
\cite{V69}. In all other cases we are in the situation of Theorem 
\ref{Theorem1}. In these cases we have a logarithmic divergence of
$\mathbb{E}(L_{\varepsilon })$ as $\varepsilon \searrow 0$, see \eqref{3Eq4}. 
Combining this moderate divergence with the rate of convergence provided in 
Theorem \ref{Theorem1}, the proof for integrability of 
\eqref{l1} for small enough non-negative $g$ follows very close along the lines
of \cite[proof of Step 3]{V69}. More precisely, by \eqref{3Eq4}, 
for $0<\varepsilon\leq1$ there exists a positive constant $k$ such that
\[
L_{\varepsilon, c}\geq -\mathbb{E}(L_{\varepsilon})\geq -k-\frac{T}{2H(2\pi)^{d/2}}|\ln(\varepsilon)|.
\]
For any constant $N\geq k+\frac{T}{2H(2\pi)^{d/2}}|\ln(\varepsilon)|$ one has
\begin{eqnarray*}
\mathbb{P}(L_c\leq -N)&=&\mathbb{P}(L_c-L_{\varepsilon, c}\leq -N-L_{\varepsilon, c})\\
&\leq&\mathbb{P}\left(|L_{\varepsilon, c}-L_c|\geq N-k-\frac{T}{2H(2\pi)^{d/2}}|\ln(\varepsilon)|\right).
\end{eqnarray*}
An application of Chebyshev's inequality then yields
\[
\mathbb{P}(L_c\leq -N)\leq \frac{\mathbb{E}(|L_{\varepsilon, c}-L_c|^2)}{\left(N-k-\frac{T}{2H(2\pi)^{d/2}}|\ln(\varepsilon)|\right)^2}\leq
K\frac{\varepsilon^{1/2}}{\left(N-k-\frac{T}{2H(2\pi)^{d/2}}|\ln(\varepsilon)|\right)^2}.
\]
In particular, for
\[
\varepsilon = \exp\left(-H(2\pi)^{d/2}(N-k)/T\right)
\]
one obtains
\[
\mathbb{P}(L_c\leq -N)\leq \frac{4K}{(N-k)^2}\exp\left(-H(2\pi)^{d/2}(N-k)/(2T)\right).
\]
Hence, there exists a positive constant $M$ such that \eqref{l1} is integrable 
for all $0 \le g \le M$.

\noindent
(ii) In the cases $dH < 1$ we know from \cite[Theorem 1 (i)]{NH05} that
the following limit exists
\begin{equation*}
0 \le L := \lim_{\varepsilon \searrow 0} L_{\varepsilon } \ \mbox{in} \ L^2.
\end{equation*}
Thus, $\exp(- gL)$ is integrable for all non-negative $g$.
\hfill $\blacksquare \medskip$

\section*{Appendix}

The following lemma is an immediate consequence of the Cauchy--Schwartz inequality.
\begin{lemma}\label{streit}
Let $0 < \alpha, \beta < \infty$ and $1/2 < m < \infty$, then there exists
a positive constant $A$ such that 
\begin{equation*}
\int_{0}^{\varepsilon }dx\,(\alpha +\beta x)^{-m} \leq A \varepsilon^{1/2}
\alpha^{-m+1/2}\beta^{-1/2}.
\end{equation*}
\end{lemma}

For $i = 2,3$ we set 
\begin{equation*}
\xi_i(x) := \frac{1}{(\delta_i +x\rho_i )^{(D+1)/2}}
-\frac{1}{((\lambda_i +x)\rho_i)^{(D+1)/2}}, \quad x\geq 0.
\end{equation*}
The following lemma is a generalization of estimates (56) and (57) obtained in 
\cite[Lemma 10]{NH05}.
\begin{lemma}\label{grothaus}
For $i = 2,3$ there exists a positive constant $B$ such that
\begin{align}
\xi_i(x) &\leq B \mu_i^2((\lambda_i+x)\rho_i)^{-(D+1)/2-1},  \label{56} \\
\xi_i(x) &\leq B \left( \left( \lambda_i +x\right) \rho_i
\right)^{-(D+1)/2}, \label{57}
\end{align}
for all $x \geq 0$.
\end{lemma}

\noindent \textbf{Proof.} Estimate (\ref{56}) implies estimate (\ref{57}).
Indeed, according to \cite[Lemma 3 (2)]{H01}, for some suitable constant 
$0<k<1$, 
\begin{equation*}
\lambda_i\rho_i-\mu_i^2=\delta_i\geq k\lambda_i\rho_i. 
\end{equation*}
Since $\lambda_i,\rho_i$ are positive, this implies that 
\begin{equation}
\mu_i^2\leq (1-k)\lambda_i\rho_i\leq (1-k)(\lambda_i+x)\rho_i,  \label{5Eq1}
\end{equation}
for all $x\geq 0$. Thus, assuming (\ref{56}), (\ref{57}) follows from 
(\ref{5Eq1}). 

Therefore, the proof amounts to prove (\ref{56}). Given 
\begin{equation}
\xi_i(x)=\left(\left(1-\frac{\mu_i^2}{(\lambda_i+x)\rho_i}
\right)^{-(D+1)/2}-1\right)((\lambda_i+x)\rho_i)^{-(D+1)/2}  \label{5Eq2}
\end{equation}
observe that due to \eqref{5Eq1} 
\begin{equation*}
0\leq \frac{\mu_i^2}{(\lambda_i+x)\rho_i}\leq 1-k<1. 
\end{equation*}
Hence let us consider the function
\begin{equation*}
[0, 1-k] \ni y \mapsto f(y):=(1-y)^{-(D+1)/2}-1 \in [0,\infty).
\end{equation*}
Since $f(0)=0$ and $f^{\prime}$ is continuous on $(0, 1-k)$ with a continuous 
continuation to $[0, 1-k]$, there exists a positive constant $B$ such that
\begin{equation*}
f(y)\leq \max_{z\in [0,1-k]}|f^{\prime}(z)| y \leq B y
\quad \mbox{ for all } \quad y \in [0,1-k]. 
\end{equation*}
Applying this inequality to (\ref{5Eq2}) yields the required 
estimate (\ref{56}).
\hfill$\blacksquare$

\begin{lemma}\label{le3}
For $i = 2,3$ there exists a positive constant $C$ such that
\begin{align}
\Xi_i^{\varepsilon} &\leq C \varepsilon^{1/2} \rho_i^{1/2} \mu_i^{2} \left(\lambda_i\rho_i\right) ^{-(D+2)/2},\label{357}\\
\Xi_i^{\varepsilon} &\leq C \varepsilon^{1/2} \rho_i^{1/2} \left( \lambda_i \rho_i \right) ^{-D/2}, \label{356}
\end{align}
for all $\varepsilon > 0$.
\end{lemma}
\noindent \textbf{Proof.} Recall that
\begin{equation*}
\Xi_i^\varepsilon = \rho_i \int_{0}^{\varepsilon }dx\,\xi_i(x), \quad i = 2,3.
\end{equation*}
Hence \eqref{357} and \eqref{356} follow from \eqref{56} and \eqref{57}, 
respectively, together with Lemma \ref{streit}.
\hfill$\blacksquare$

\subsection*{Acknowledgments}

Financial support by PTDC/MAT/67965/2006, PTDC/MAT/100983/2008 and
FCT, POCTI-219, ISFL-1-209 and the hospitality of the Physics Dept., MSU-IIT, 
Iligan, the Philippines, are gratefully acknowledged.


\begin{thebibliography}{GRV03}

\bibitem[BC95]{Biswas}
P.~Biswas and B.~J. Cherayil.
\newblock Dynamics of fractional {B}rownian walks.
\newblock {\em J. Phys. Chem.}, 99:816--821, 1995.

\bibitem[Bol93]{Bolthausen}
E.~Bolthausen.
\newblock On the construction of the three dimensional polymer measure.
\newblock {\em Probab. Theory Related Fields}, 97:81--101, 1993.

\bibitem[CY87]{CY87}
J.~Y. Calais and M.~Yor.
\newblock Renormalisation et convergence en loi pour certaines int\'egrales
  multiples associ\'ees au mouvement brownien dans $\mathbb{R}^d$.
\newblock {\em Lecture Notes in Math.}, 1247:375--403, 1987.

\bibitem[Edw65]{E65}
S.~F. Edwards.
\newblock The statistical mechanics of polymers with excluded volume.
\newblock {\em Proc. Phys. Sci.}, 85:613--624, 1965.

\bibitem[GRV03]{GRV03}
M.~Gradinaru, F.~Russo, and P.~Vallois.
\newblock Generalized covariations, local time and {S}tratonovich {I}t\^o's
  formula for fractional {B}rownian motion with {H}urst index {$H\ge\frac14$}.
\newblock {\em Ann.~Probab.}, 31(4):1772--1820, 2003.

\bibitem[HN05]{NH05}
Y.~Hu and D.~Nualart.
\newblock Renormalized self-intersection local time for fractional {B}rownian
  motion.
\newblock {\em Ann.~Probab.}, 33:948--983, 2005.

\bibitem[HN07]{NH07}
Y.~Hu and D.~Nualart.
\newblock Regularity of renormalized self-intersection local time for
  fractional {B}rownian motion.
\newblock {\em Commun.~Inf.~Syst.}, 7(1):21--30, 2007.

\bibitem[HNS08]{HNS08}
Y.~Hu, D.~Nualart, and J.~Song.
\newblock Integral representation of renormalized self-intersection local
  times.
\newblock {\em J. Funct. Anal.}, 255:2507--2532, 2008.

\bibitem[Hu01]{H01}
Y.~Hu.
\newblock Self-intersection local time of fractional {B}rownian motions - via
  chaos expansion.
\newblock {\em J. Math. Kyoto Univ.}, 41:233--250, 2001.

\bibitem[NOL07]{NL07}
D.~Nualart and S.~Ortiz-Latorre.
\newblock Intersection local time for two independent fractional {B}rownian
  motions.
\newblock {\em J.~Theoret.~Probab.}, 20(4):759--767, 2007.

\bibitem[OSS10]{OSS09}
M.~J. Oliveira, J.~L. Silva, and L.~Streit.
\newblock Intersection local times of independent fractional {B}rownian motions
  as generalized white noise functionals.
\newblock Acta Appl. Math., doi:10.1007/s10440-010-9579-1 (published online),
  2010.

\bibitem[Ros87]{R87}
J.~Rosen.
\newblock The intersection local time of fractional {B}rownian motion in the
  plane.
\newblock {\em J.~Multivar.~Anal.}, 23:37--46, 1987.

\bibitem[Sim74]{Si74}
B.~Simon.
\newblock {\em The $P(\phi)_2 $ {E}uclidean (Quantum) Field Theory}.
\newblock Princeton University Press, Princeton, New Jersey, 1974.

\bibitem[Var69]{V69}
S.~R.~S. Varadhan.
\newblock Appendix to ``{E}uclidean quantum field theory" by {K}.~{S}ymanzik.
\newblock In R.~Jost, editor, {\em Local Quantum Theory}, New York, 1969.
  Academic Press.

\bibitem[Wes80]{W80}
J.~Westwater.
\newblock On {E}dwards' model for long polymer chains.
\newblock {\em Comm. Math. Phys.}, 72:131--174, 1980.

\bibitem[Yor85]{Y85}
M.~Yor.
\newblock Renormalisation et convergence en loi pour les temps locaux
  d'intersection du mouvement brownien dans $\mathbb{R}^3$.
\newblock {\em Lectures Notes in Math.}, 1123:350--365, 1985.

\end{thebibliography}

\end{document}